# A MKID-readout based on a heterogeneous, closely coupled architecture

Gerrit Grutzeck[*,1], Ingo Krämer[1], Miroslaw Ciechanowicz[1], Nicolas Reyes[1], Carsten König[1],
Andrey Baryshev[2], Stephen Yates[3] and Bernd Klein[1]

*Abstract*—Within this proceeding, we introduce the U-Board platform, a versatile platform for signal generation, acquisition and processing, based on a heterogenous processing architecture. Based on this platform we present a readout for Microwave Kinetic Inductance Detectors (MKIDs) for the A-MKID camera for APEX. In addition to the implementation of the readout on this heterogenous architecture, we also present a first comparison of the performance of the readout compared to the currently used readout of the A-MKID camera. Last but not least, we discuss how we plan to miniaturize the current prototype, which is based on commercial off the shelf components.

## I. INTRODUCTION

Microwave Kinetic Inductance Detectors (MKIDs) have a great potential for large and sensitive detector arrays for use in sub-mm wavelength imaging and mid-resolution spectroscopy [1]. As these detectors are superconducting, notchfilters with a high-quality factor, they can be easily multiplexed in the frequency domain to allow for detector arrays with a high number of detectors [2]. Furthermore, the fabrication process is based on conventional optical or electron beam lithography, which leads to simple and reliable production. Also, the detectors can achieve photon noise limited sensitivity performance [3]. Due to the break-up of Cooper-Pairs by photons in the superconductor, the resonance frequency of MKIDs is a function of the photon-flux onto the detector.

To read out such detectors, the phase and amplitude changes of a readout tone placed close to the resonance frequency of the detector can be used. For this readout technique, the backend has to generate a tone for each detector, close to the detector's resonance frequency. The amplitudes and phases of these tones are then altered due to the interactions with the notch-filters of the detectors. At the end, each tone is analyzed again by the backend and the photon flux is determined based on the changes of signal, especially based on the phase change of the signal. Therefore, a backend for reading out MKIDs needs a phase coherent signal generation and acquisition combined with the processing power needed to generate and analyze the readout signal. The U-Board platform combines the strengths of General Purpose Graphical Processors (GPGPUs) and Field Programmable Gate Arrays (FPGAs), together with a phase stable signal generation and acquisition in the analog frontend. This platform is optimally suited to implementing a MKID readout.

The A-MKID instrument is a dual-color incoherent camera for the APEX telescope [4], which offers excellent observation conditions up to 300 μm based on MKIDs. The high frequency array is sensitive to 350 μm and the low frequency array is sensitive to 870 μm [5]. Both arrays are co-aligned and observe the same area of 15 arcmin x 15 arcmin of the sky. The nominal resonance frequencies of the detectors lie between 4 GHz and 8 GHz. In total the camera has 24 readout chains, where each chain hosts between 680 to 880 MKIDs.

The readout presented here is optimized for the A-MKID camera. Therefore, it aims for an analog bandwidth of 4 GHz for up to 1280 tones. The processing is based on a one million channel Fast Fourier Transformation (FFT), which is split between the FPGA and GPGPU. Furthermore, additional processing and reduction steps are migrated to the GPGPU to reduce the work-load on the controlling computer. It should replace the currently used readout of the A-MKID camera, which uses a heterodyne mixing scheme and lower resolution converters and therefore limits the performance of the camera.

## II. THE U-BOARD PLATTFORM

In recent times, the performance of Digital-Analog-Converters (DACs) and Analog-Digital-Converters (ADCs) have not only improved strongly, but also the processing power of FPGAs and GPGPUs has increased a lot. This opens up new possibilities to build wideband readouts for MKID cameras, which can handle a thousand detectors per readout chain. Furthermore, the development of GPGPU modules for embedded systems, has removed the need for dedicated computers for hosting the GPGPU. This allows for compact heterogeneous architectures based on FPGAs and GPGPU modules. In the following, we will discuss the U-Board platform based on commercial off the shelf components as shown in Fig. 1, with an analog frontend optimized for the readout of the A-MKID camera.

### A. Analog frontend and IF processing

To reduce the phase noise between a number of local oscillators (LOs), a homodyne mixing scheme is implemented. Therefore, the same LO is used for up and down conversion of the baseband signals. For this up and down converting IQ-mixers are used to reduce the required bandwidth of the DACs to 2.1 GHz, which allows the selection of 16 bit DACs. Furthermore, the IQ-mixers allow for a digital optimization of the sideband separation. The concept of the analog frontend and the IF processing is shown in the left part of Fig. 3.

For the signal generation, two DACs of the type AD9174 are used, each of which can generate a signal with a resolution of 16 bit with an instantaneous bandwidth of up to 3.08 GHz. The signal acquisition is done with a single ADC12DJ5200RF, which can sample an instantaneous bandwidth of up to 5.2 GHz with a resolution of 12 bit. The ADC is used in dual channel mode, where each channel has an instantaneous bandwidth of up to 2.6 GHz and typically an effective number of bits of 8.7.

[1]Max Planck Institute for Radio-Astronomy, Auf dem Hügel 69, 53121 Bonn, [2]Kapetyn Astronomical Institute, University of Groningen, The Netherlands, [3]SRON, Netherland Institute for Space Research, Groningen, The Nehterlands

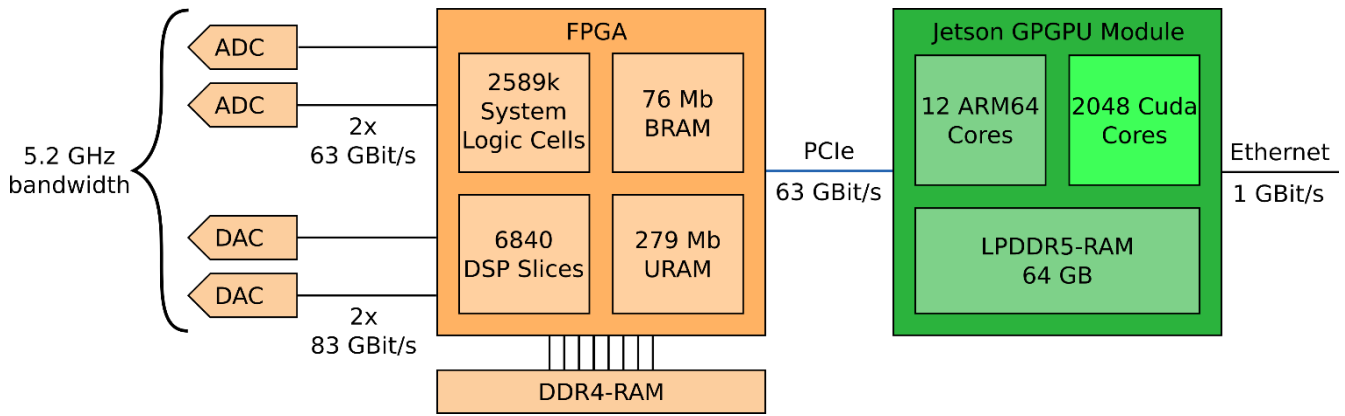

Fig. 1. The architecture of the U-Board prototype based on commercial off the shelf hardware. The DACs and ADCs are hosted on a custom PCB, which is connected to FPGA with high speed serial links (JESD204C). The FPGA (UltraScale+ Virtex 9) is from the high performance series of UltraScale+ FPGAs. Additional to the internal memory of the FPGA two 80-bit wide DDR4 memories are associated with the FPGA. The FPGA is finally connected to the Jetson GPGPU Module via a PCIe connection.

*B. Heterogeneous architecture*

Both FPGAs and GPGPUs have their own strengths and weaknesses. FPGAs are per definition hard real-time capable and have flexible connection options. The combination of hard real-time capabilities and flexible interfacing options make possible to connect high-speed DACs and ADCs electrically and protocol wise. Furthermore, FPGAs have good integer processing power. But, their floating-point processing power and memory size are limited. Nether the less, the onboard memory of FPGAs has a bandwidth with very low latencies. GPGPUs on the other hand, do not have hard real-time capabilities, but need carefully designed buffers to ensure soft real-time capabilities. This behavior, is due to the fact, that GPGPUs can only be used in combination with an operating system. But, compared to FPGAs, GPGPUs deliver a higher floating-point processing power and have huge memory space. These memories also have a high bandwidth (not as high as the memory on FPGAs), but the latency of the memory is much higher. Last but not least, the GPGPU have a programming model, which is much closer to conventional software programming compared to the hardware description language used for FPGAs. This allows for faster implementations of new algorithms and ideas.

The idea of a heterogeneous architecture, based on FPGAs and GPGPUs, is to combine the strengths of both architectures. Two possible interconnections can be used to connect FPGAs and GPGPUs; high speed Ethernet (e.g. 100 Gbit/s) and PCIe. Ethernet can be set up for multicast applications, where one source sends data to multiple sinks, while PCIe is used for more point-to-point like connections. The disadvantage of high-speed Ethernet is, that the infrastructure costs (e.g. fiber links, switches, network cards) is significant. Therefore, high speed Ethernet has advantages only for applications, which rely on multicast data transfers (e.g. beam former). In heterogeneous architecture based on a PCIe interconnect, the FPGA and GPGPU are typically hosted in a powerful computer, as neither the FPGA nor the GPGPU are fully suited as PCIe host. This approach is again raising additional costs due to the requirements to host a FPGA and a GPGPU in a computer. Furthermore, a computer per FPGA is not compatible with the space requirements of typical backends, as the space close to the instrument is limited. An alternative to using a conventional computer to host both, the FPGA and the GPGPU, are the Jetson modules produced by Nvidia. These systems on a module include a multicore ARM CPU, a high performance GPGPU and a huge, high bandwidth memory region. These Jetson modules can be used as a PCIe host to which the FPGA is connected. This allows for a simple and fast point to point connection between the FPGA and the GPGPU.

To speed up the development time, we first developed a prototype mostly based on commercial off the shelf components, which is shown in Fig. 2. Only the analog frontend with the DACs and ADCs is hosted on a custom printed circuit board. This analog frontend is connected with a high-speed connector (FMC+) to the evaluation board VCU118 from Xilinx. From the hardware present on this evaluation board, we utilize FPGA, an UltraScale+ Virtex 9, as well as one of the DDR4-Memory banks for the MKID readout. The FPGA evaluation board is then connected to a development Kit from Nvidia with 8 lanes of PCIe 3. The development Kit from Nvidia is built around the Jetson Orin AGX module, which supplies the platform with 2048 CUDA cores (the GPGPU part of the module), 32GB of memory and 12 CPU cores, for housekeeping and controlling.

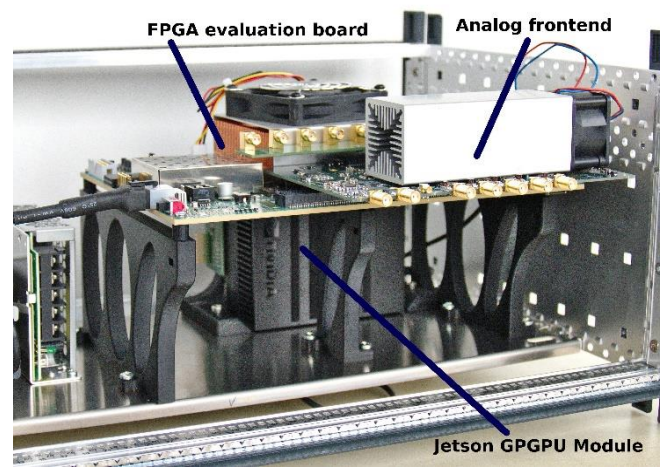

Fig. 2. The Picture of the prototype based on the commercial off the shelf hardware. In the foreground, the analog frontend, based on a custom PCB, is visible. This frontend is connected with a high-speed connector (FMC+) to the evaluation board VCU118 from Xilinx, which hosts the FPGA and the associated DDR4 memory. In the background, the GPGPU development Kit from Nvidia, with the Jetson GPGPU Module, is visible. This module is connected to the FPGA via PCIe.

### III. READOUT

In the following we will discuss how the readout is implemented based on the platform described above, and then give a preliminary comparison to the currently used readout for the A-MKID camera.

#### A. Implementation of the readout

The readout is based on a one million point FFT, which is split into a first stage on the FPGA and a second stage on the GPGPU. Therefore, the possible frequencies of the readout tones are limited to a finite grid with an approximate spacing of 4 kHz. The concept of the readout and its division onto the FPGA and GPGPU is shown in the right part of Fig. 3.

The waveform for the signal generation via the DACs is streamed via the FPGA from the DDR4-memory to the DACs. While setting up the readout, this waveform is generated on the GPGPU based on the determined positions of the MKIDs. The generation of the waveform also takes into account a frequency dependent calibration of signal path from the DACs to the IQ up converter, to compensate for differences between the I and Q paths, as well as non-perfection in the IQ-mixer itself. Furthermore, blindtones are added to the waveform, which are used to correct for amplitude and phases changes due to changes of the system (e.g. drifting reference voltages, changes in propagation delay).

The real-time analysis of the digitized signal is split between the FPGA and the GPGPU. The first processing step is an 1M ($2^{20}$) point FFT. This FFT is split between the FPGA and the GPGPU, as described in the following. On the FPGA, 16 parallel FFTs of size of 64k are calculated in fixed point (see 16x 64k FFT in Fig. 3). Then the frequency bins of this FFTs, which contain the tones, are selected and transferred to the GPGPU via PCIe (see Channel Mask in Fig. 3). This reduces the data rate, which has to be transferred via PCIe, and further reduces the workload on the GPGPU. On the GPGPU the results from the 16 parallel FFTs are combined into the result of the 1M point FFT with a modified Discrete Fourier Transformation, which takes the calibration for the sideband separation of the down converter into account. This result corresponds to the complex S21 value of the system. The results of the consecutively calculated FFTs are interpreted as time-data with a sample rate of $\frac{4\,\text{GHz}}{2^{20}} = 4.01$ kHz, which carry the changes of the photon flux on the detectors. In the next step, the data rate of this time-data is further reduced with a configurable filter (configurable length and coefficients) and a down sampling stage. The down sampling stage allows for an integer down sampling ratio; therefore, the possible sampling rates are $\frac{4.01\,\text{kHz}}{n}$ with $n \geq 1$.

To reconstruct the phase change of the readout tone with respect to the MKID resonance, the FFT channels are first divided by the blindtones. This enhances the stability of the system as common, unwanted effects (e.g. drift of voltage references, changes of the propagation delay) are compensated for. In the next step, the calibration of the MKID is applied to map the complex S21 to the phase change with respect to the resonance. This phase change can be used to extract the frequency shift of the resonance, especially as the relationship is linear for small frequency shifts, which is the operation point of the A-MKID camera. All these reconstruction steps are implemented on the GPGPU. Finally, the data is streamed out of the Jetson Module to the controlling computer.

The readouts for one camera color are controlled by a central software instance on a computer (as shown in Fig. 5). All readout chains are connected via a high-speed Ethernet (e.g. 40 Gbit/s) to the controlling computer. Over this connection the chains are controlled, and the reduced data is sent from the chain to the computer, where all data streams are merged. The merged data stream is then sent out via another high-speed Ethernet connection to the telescope infrastructure.

#### B. Performance of the readout

In the following, we will briefly compare the performance of the readout based on the U-Board platform with the performance of the currently used readout for the A-MKID camera. The currently used readout uses a more complicated heterodyne mixing approach. Furthermore, the resolution of the DACs and ADCs is smaller. Particularly, the ADC only has 7.6 effective number of bits. Therefore, we expect a better performance of the presented readout compared to the readout currently under use.

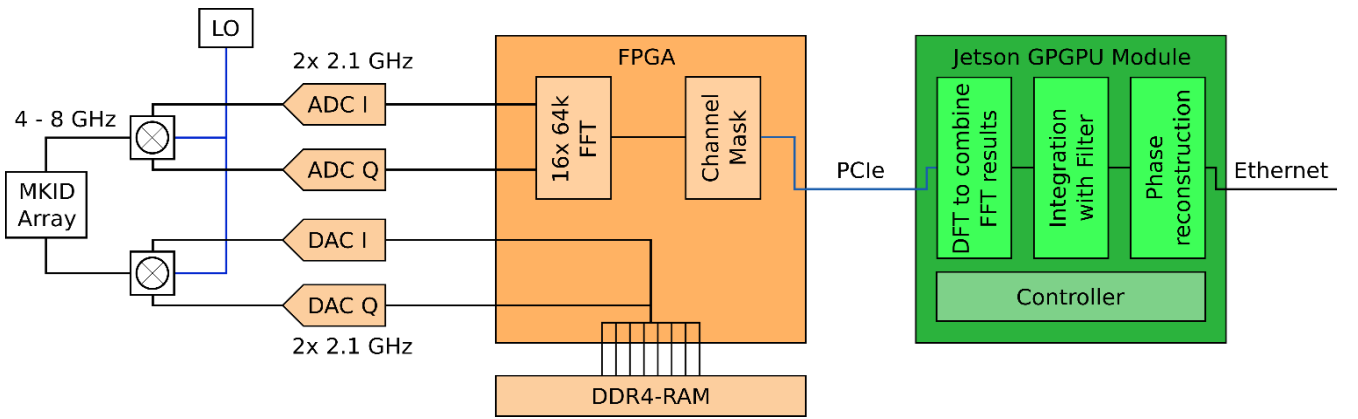

Fig. 3. The concept of the implemented MKID readout based on the U-Board for a single chain. The IF processing and analog frontend uses a homodyne mixing schema based on IQ-mixers to deliver 4GHz of usable, instantaneous bandwidth for reading out the MKIDs in the detector array. The signal with the readout tones is streamed from the DDR4-Memory associated with the FPGA. For the signal analysis, the first step is a one million point FFT split between the FPGA and GPGPU. Here only the spectral channels, which include a tone, are transferred and processed on the GPGPU. To reduce the data rate the complex results of the FFT are downsampled with a configurable antialiasing filter. Finally, the phase change due to the shifting of the MKIDs is extracted and send out via Ethernet.

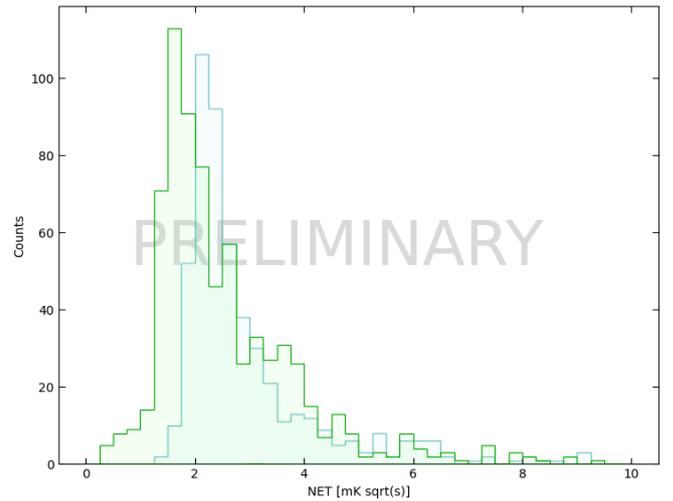

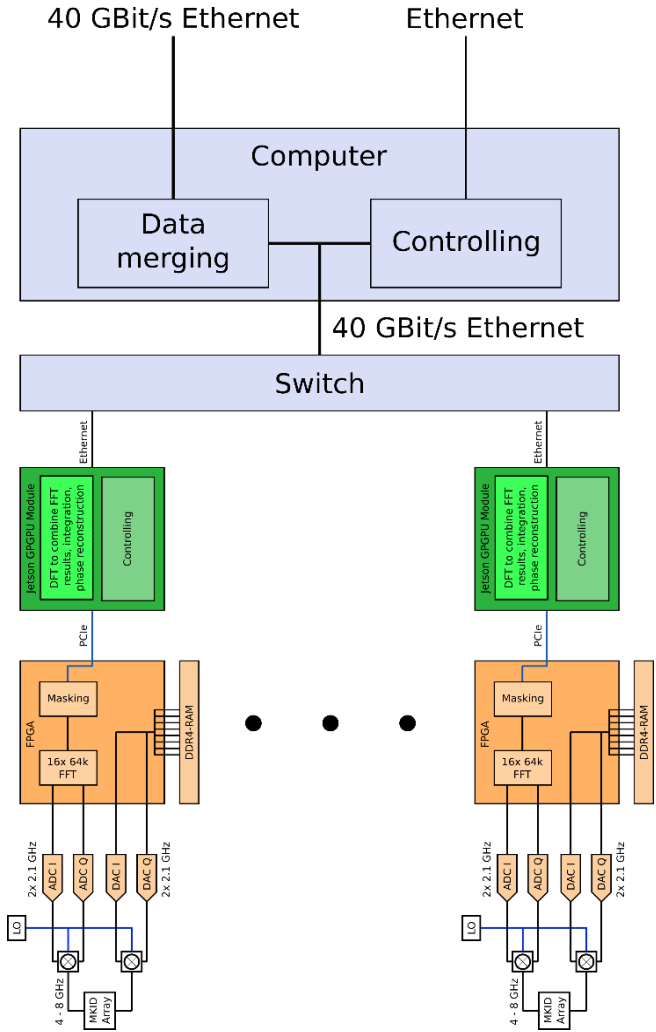

Fig. 5. The concept to combine a number of readout chains into a single instrument. For instruments with more than a single chain, like the A-MKID camera, the controlling of all chains and the data from all chains has to be centralized. As all the online reduction of the data is implemented on the heterogeneous U-Board, the data coming from the readout chains only has to be merged into a single data stream, which is send out to be saved. This reduced the requirements on the controlling computer enormously, as no demanding data reduction has to be done on it.

In order to compare the performance of the two readout systems, measurements with the identical chain of detectors and the identical optical path are executed. The performance is measured with a noise scan, which is calibrated with a wirescan. For the noise scan, the camera is pointing at a bucket with liquid nitrogen, which is used to simulate the sky at APEX. The wirescanner moves two cables (one in x- and on in y-direction) in front of the liquid nitrogen through the field of view of the camera. These cables generate a signal of reproducible and known strength, which is used to calibrate the noise amplitude in the noise scan. As these measurements are done with a non-final optical filter stack and under laboratory conditions, the results are only preliminary.

In Fig. 4 the measured histograms, one for each readout,

Fig. 4. Comparison of the sensitivities of the current readout with the sensitivities measured with the readout based on the U-Board platform. The green histogram represents the measured sensitivities with the readout based on the U-Board platform, and the blue one represents the sensitivities measured with the currently used readout.

of the sensitivities of one chain of the camera are shown. The green histogram represents the readout based on the U-Board platform, as presented here. The blue histogram shows the performance of the currently used readout. Two important aspects are visible in this figure. Firstly, the readout based on the U-Board platform gives a better sensitivity. This is expected, as the currently used readout is limiting the camera performance and the U-Board based readout uses ADCs with better performance. Secondly, the yield of MKIDs is higher. This is one of the benefits of using a side band separating, homodyne mixing scheme instead of a non-sideband separating, heterodyne mixing scheme. The separation of side bands reduces the number of MKIDs, which are close to each other in the baseband. These close MKIDs are not suitable for operation, as the crosstalk between them is significant.

IV. OUTLOOK AND CONCLUSION

We have presented a MKID readout, based on a 1M FFT, with a bandwidth between 4 GHz and 8 GHz, which is implemented on the U-Board platform, a heterogeneous architecture. The various processing architectures in the U-Board allow for a complete implementation of the online reduction pipeline on the backend. This reduces the requirements regarding the central, controlling computer. The choice of an FFT based approach, with 1M points, allows for a high maximal sample rate of 4 kHz and a fine grid of possible readout tone positions (also 4 kHz). Furthermore, the prototype based on the U-Board architecture for the A-MKID readout is performing better than the currently used readout, both in terms of sensitivities and in terms of yield. But the final performance of the readout compared to the current one has to be demonstrated on the sky.

As the U-Board platform is flexible, not only can a MKID readout be implemented on it, but also other types of backends, a high-resolution spectrometer for example.

The prototype based on commercial off the shelf components has no compact form, we are developing a more integrated solution. This new solution is based on a complete custom printed circuit board, which includes the analog frontend, an UltraScale+ Kintex 15 FPGA, the DDR4-memory associated with the FPGA and the connector for the Jetson module. As the analog frontend and the signal processing is the same on the new solution, as on the commercial off the shelf components prototype, we expect the same analog performance, with less space and power consumption.